\documentclass[preprint,12pt]{elsarticle}

\usepackage{amssymb}

\usepackage{makecell}
\usepackage{url}

\journal{Information Systems}

\begin{document}

\begin{frontmatter}



\title{Process-Driven Visual Analysis of Cybersecurity Capture the Flag Exercises}

\author[inst1]{Radek O\v{s}lej\v{s}ek}
\author[inst1]{Radoslav Chudovsk\'{y}}
\author[inst1]{Martin Macak}

\affiliation[inst1]{organization={Masaryk~University,~Faculty~of~Informatics},
            addressline={Botanicka~68a}, 
            city={Brno},
            postcode={60200}, 
            country={Czech~Republic}}

\begin{abstract}
Hands-on training sessions become a standard way to develop and increase knowledge in cybersecurity. As practical cybersecurity exercises are strongly process-oriented with knowledge-intensive processes, process mining techniques and models can help enhance learning analytics tools. The design of our open-source analytical dashboard is backed by guidelines for visualizing multivariate networks complemented with temporal views and clustering. The design aligns with the requirements for post-training analysis of a special subset of cybersecurity exercises -- supervised Capture the Flag games. Usability is demonstrated in a case study using trainees' engagement measurement to reveal potential flaws in training design or organization.
\end{abstract}


\begin{highlights}
\item Practical cybersecurity exercises are strongly process-oriented
\item They are characterized by knowledge-intensive processes
\item Process graphs that can provide insight into trainees' behavior and possible exercise suffer from complexity
\item Rules for the visualization of multivariate networks are used for generic design decisions
\item A sliding window algorithm is used to analyze changes in trainees' behavior
\item Clustering techniques are used to classify individual and collective behavior
\end{highlights}


\end{frontmatter}



\section{Introduction} \label{sec:introduction}

Hands-on training sessions have become a standard way to develop and increase knowledge in cybersecurity. They are often organized in cyber ranges~\cite{yamin2020} that provide online environments with data monitoring. These environments typically consist of virtualized computers mutually connected into an emulated network. They are transparently accessible to trainees as if they were physical devices. Learning goals are typically related to attack or defense skills, e.g., finding the vulnerabilities in the network or protecting the infrastructure from attackers. Moreover, the cyber ranges usually provide some content management system, which is used to inform trainees about expected cybersecurity tasks and to supervise and coordinate training sessions.

\subsection{Process Mining Background}

Practical cybersecurity exercises are strongly process-oriented. Captured data logs encode basic steps conducted by trainees in solving cybersecurity tasks. Even though these event logs represent the primary source of information for learning analytics, they are too low-level and difficult to analyze. 

Activity graphs produced by process mining techniques on top of the raw event logs provide behavioral models of higher abstraction. Analysts could use them to gain insight into behavioral patterns from which hypotheses about students' progress, difficulties, or results can be formulated and further investigated~\cite{PMinKYPO2}. However, a suitable integration of process mining into the learning analytics of cybersecurity exercises remains an open question.

This paper explores how visual interactions, complementary to standard process graphs, can support training designers and supervisors in analyzing post-training data within a process-driven analytical framework.

\subsection{Cybersecurity Training Background} \label{sec:cyber-training-background}

The cybersecurity exercise playthroughs can significantly differ based on the participants' knowledge, which is one of the characteristics of knowledge-intensive processes~\cite{gronau2005kmdl}. A common approach to addressing this challenge is to reduce the complexity of the analyzed process~\cite{vidgof2023impactprocesscomplexityprocess}. We examined proposed measures for quantifying process complexity and chose to focus on reducing process average affinity -- a complexity measure that calculates the mean of the pair-wise relative overlap of direct following relations over all sequences in the event log~\cite{gunther2009process}. 

In this scenario, variation measures cannot be reduced due to the flexibility participants have in solving cybersecurity exercises, which often results in each trace representing a unique process variant. Theoretically, size measures could be reduced through specialized exercise assignments that limit the number of steps a participant can take (e.g., restricting them to three commands to succeed, thereby shortening the process sequence) or by limiting the player in a very closed environment without much freedom (e.g., providing only a web page with three buttons to use, thereby reducing the number of events). However, such constraints are uncommon in cybersecurity exercises~\cite{ISMAIL2024100501} and fall outside the scope of this paper, as exercise assignment design occurs before post-training analysis.

We restrict the application domain to supervised training. This type of education represents an obvious approach to cybersecurity learning, where multiple training sessions are organized for a limited group of students under the supervision of a tutor. Unlike courses that can be attended at any time without restrictions, e.g., online exercises, supervised lessons take place under controlled conditions. This results in a lower process average affinity, where the data -- playthroughs from multiple trainees -- are less complex from this perspective and more suitable for joint behavioral analysis. 

Nonetheless, even restricting ourselves to supervised sessions, the cybersecurity training still proposes high variability of learning principles ranging from table-top-like exercises~\cite{angafor2020,vykopal2024}, jeopardy games, or attacker-defender competitions to capture the flag (CTF) games~\cite{svabensky2021,zennaro2023}.  Cyber-defense exercises (CDX)~\cite{seker2018,almroth2020,maennel2023} represent a most advanced type of realistic training that mimics real-world scenarios involving multiple cooperating and enemy teams. Individual concepts can significantly differ in the freedom of achieving the learning objective and then in the average affinity of process models.

We have analyzed the process complexity of all these exercise types and selected CTF games as the ones with the lowest process average affinity. Compared to, e.g., CDXs, the training scenarios are more specific, and players' behavior is less variable. Tasks are not solved ad-hoc. We can suppose that participants will take similar steps at similar times. Moreover, in many cases, CTF games are divided into multiple levels, allowing the process to be split into subprocesses, thus reducing the process complexity through the process size. Therefore, this paper primarily focuses on supervised CTF games.

Cybersecurity CTF games belong to so-called puzzle-based learning, which represents a well-known gamification principle where puzzles are used as a metaphor for getting students to think about how to frame and solve unstructured problems~\cite{michael2005,michalewicz2007}. Typical puzzles in cybersecurity learning are, for example, techniques of scanning the network and revealing open services or techniques of privilege escalation. An example of the gamification principles of the jeopardy CTF and CDX can be found in~\cite{russo2023}. 

From the perspective of process modeling, puzzle-based content prescribes consequential tasks as a sequence of expected actions leading to a milestone -- finding the task's solution. For example, the \emph{ipconfig}, \emph{nmap}, or \emph{traceroute} commands can be run with multiple parameters on a trainee's computer connected to the network to explore the network state during the reconnaissance phase of the training scenario. Obtained event logs -- commands and actions performed by trainees, provide a kind of basic logical structuring that could be used for drill-down exploration and tackling the information complexity of the whole training session.

\subsection{Goals and Research Questions}

Despite restricting the application domain to learning strategies that reduce the process average affinity, the application of process mining for their analysis remains challenging. Furthermore, the analysis of commands used to solve cybersecurity tasks brings additional obstacles to process mining utilization~\cite{PMinKYPO2}. We aim to overcome these challenges by using visual analytics methods, where the process models represent multivariate networks that can serve as higher abstraction models for exploratory visual analysis. In this paper, we formulate the following research questions:
\begin{enumerate}
    \item[RQ1] \emph{What are the properties of process graphs derived from event logs of supervised CTF games?} The complexity of process graphs influences the design of visualization techniques used to explore them from different perspectives. Understanding the properties of process graphs obtained from real data is, therefore, essential for developing meaningful visual interactions.
    \item[RQ2] \emph{What are the objectives of learning analytics that process mining can help overcome?} Post-training analysts formulate a wide range of analytical questions and hypotheses. Our goal is to identify objectives that are frequent, meaningful, and effectively addressable through process-driven exploration.
    \item[RQ3] \emph{What design principles should guide the development of a process-driven analytical dashboard?} There is no universal solution for analyzing cybersecurity training data and meeting diverse analytical objectives. Thus, we seek to outline key principles for designing tailored process-driven analytical dashboards.
\end{enumerate}

\section{Methodology}

Our approach to using process mining in the visual analytics of CTF games is backed by the conceptual model for the visual analysis process~\cite{sacha2014knowledge}. This framework is characterized by the interaction between data, data models, visualizations, and hypotheses-based knowledge discovery of analysts. 

Data collected from CTF games consists of raw event logs that sparsely capture trainees' behavior. Process-oriented data models derived from these events take the form of process graphs, which are typically visualized as various types of oriented graphs. However, such static representations are insufficient for learning analytics in cybersecurity learning, as the analytical objectives of training designers and supervisors are highly variable and complex. These objectives often involve analyzing temporal changes in behavior or classifying trainees based on the tactics used to achieve cybersecurity tasks. Such objectives are difficult to accomplish using static behavioral graphs alone.

To address these diverse analytical needs, we treat process graphs as multivariate networks -- a concept already researched by the visual analytics community. Our approach extends static process graphs with complementary data models and juxtaposed visualizations that incorporate temporal and clustering aspects of data exploration while adhering to the principles and guidelines of multivariate network visualizations~\cite{nobre2019}.

Although the visualization techniques of multivariate networks provide general rules and best practices for effective exploratory interactions, designing a tailored solution remains challenging. Therefore, we use these guidelines within the \emph{Nested Model}~\cite{meyer2012four} design methodology as follows.

\paragraph{Domain problem and data characterization}
We benefited from the authors' long-term experience in organizing hands-on cybersecurity training sessions and conducting process mining analysis in various application domains. This expertise enabled us to collect raw data, conduct a preliminary analysis, and then restrict the learning domain to the supervised puzzle-based CTF games that best fit the requirements of practically usable process mining, as described in Section~\ref{sec:introduction}. Furthermore, we performed statistical analysis of raw event logs captured from CTF games and corresponding process models to reveal the basic characteristics from the perspective of multivariate networks, as discussed in Section~\ref{sec:rq1}. The results of this analysis answer the research question \emph{RQ1} and affect consequent design decisions.

\paragraph{Operation and data type abstraction}
This phase aims to map the input problems into a more specific description and then to answer the research question \emph{RQ2}. We have performed a comprehensive analysis and formulated a set of analytical questions that can benefit from the process mining usage. Then, we consulted them with domain experts -- cybersecurity training practitioners. The resulting learning objectives of tutors are discussed in Section~\ref{sec:rq2-objectives}.

\paragraph{Visual encoding and interaction design}
A typical visual representation of process mining models is a graph. We use design rules defined for the visualization of multivariate networks to extend graph-based views with complementary interactive data views. In particular, we employ juxtaposed views of temporal and clustering models, data filtering tools, and in-depth previews of activities to enhance the exploratory verification of hypotheses. Interactions and design decisions are discussed in Section~\ref{sec:rq3-solution}, answering the research question \emph{RQ3}.

\paragraph{Algorithm design}
Algorithm design aims to implement visual encoding.
We combine known principles, especially a sliding window algorithm and k-means clustering, to recognize, classify, and visualize changes in trainees' behavior. The dashboard is implemented as a functional prototype. We provide a case study for the engagement analysis in Section~\ref{sec:case-study} to validate our design.

\section{Related Work} \label{sec:related-work}

\subsection{Visual Analytics in Cybersecurity Education}

A systematic literature review of using visual analytics in education can be found in~\cite{vieira2018}. According to the authors, only a few research studies took place in classroom settings, missing an opportunity to take advantage of a controlled environment to get coherent data. This survey also reveals that the dominant visualization strategy in generic educational learning analytics is statistical views, followed by graphs. 

Cybersecurity training, where users interact with a cyber range via its interface, can be considered online learning. In \cite{kui2022}, Kyi et al. focus on the usage of visual analysis in online education. They classify visual analytics techniques into four categories. Two of them, problem-solving and learning content analysis, share the same goals with our approach. According to the authors, the problem-solving category dominantly employs glyph-embedded Sankey diagrams, statistical charts, and map charts to analyze students' cognitive and noncognitive patterns in the process of solving a series of students' problems over time~\cite{xia2020,tsung2022}. On the contrary, learning content analysis often uses node-link diagrams, keyword clouds, and concept maps as popular visual techniques to enhance the quality of online courses~\cite{huang2017,liu2018,wang2020}. However, the courses discussed in the paper are strongly based on video records. Our solution aims to utilize structured event logs.

A few papers address the application of VA methods directly in the cybersecurity education subdomain. A basic overview of analytical tasks and corresponding visual techniques can be found in~\cite{oslejsek2021}. According to their classification, this paper addresses the \emph{quality of training exercise} and \emph{behavior analysis} post-training analytical objectives. The same authors published a series of papers addressing specific solutions for visual-based feedback to trainees~\cite{vykopal2018timely,svabensky2021}, situational awareness of tutors during training sessions~\cite{burska2021enhancing}, or post-training analysis~\cite{burska2022data}. However, none of these techniques utilize process mining for behavioral analysis. 

\subsection{Educational Process Mining}

Process mining has already been applied in many educational use cases~\cite{RWBogarin2018education}. The prominent area is e-learning analysis of processes, like learning habits~\cite{rwMukala} and general usage of online platforms like Moodle~\cite{rw2Romero,Dolak2019}. Moreover, process mining was applied in other areas like curriculum mining~\cite{rwTrcka} and student registrations~\cite{rwAnuwatvisit}. Process mining potential usage was also mentioned in course achievement analysis~\cite{macakDiscord}. However, all these processes have lower process complexity than the cybersecurity exercise playthrough. 

In the case of the Git process analysis in the software development course~\cite{macak2021using}, this issue was solved by the categorization of events into smaller categories, e.g., by the size of the Git commit. However, in the case of the cybersecurity training session analysis, such categorization would end up with the loss of relevant information.

Furthermore, in the process analysis of learning difficulties~\cite{rwBannert}, a coding approach was used on the collected think-aloud data. This is a great approach to natural language processing. However, its effectiveness is limited in the process analysis of cybersecurity training session data.

Therefore, there is currently no educational process mining solution that would solve the issue of process mining applications in cybersecurity training session analysis.

\subsection{Visual Analytics in Process Mining}

Some authors have already studied visualization approaches to process mining models, regardless of the application domain. The earlier works discuss generic interaction techniques, challenges, and opportunities~\cite{kriglstein2016,gschwandtner2017}. 
Also, specific complementary visualizations for process graphs have been proposed. For instance, histograms of frequency or performance views can help to cope with deviation or bottleneck analysis~\cite{dixit2017}.

When looking for a general approach, process graphs can be considered multivariate networks. The framework proposed by Nobre et al.~\cite{nobre2019} can help conceptualize complementary visualizations that can extend traditional graph-based views on process data. Still, the framework provides a wide variety of possible solutions depending on data characteristics and the aims of analysts. Nevertheless, we use this framework as a guideline for answering the research question \emph{RQ1}.

Multivariate networks provide a static view of behavioral patterns, where time is captured by only ordering activities (edges in the process graphs). Therefore, an analytical dashboard should also adopt some temporal visualizations of event sequences~\cite{guo2021} to cope with the dynamic aspects of trainees' behavior. In this case, techniques based on sliding windows are often used for continuous query processing of spatiotemporal data streams, aiming to recognize activity patterns of human beings~\cite{ortiz2011dynamic}, discover emerging trends from temporal data~\cite{khan2010sliding}, or detect anomalies in behavior~\cite{talagala2020anomaly}. We use a sliding window approach to recognize changes in trainees' engagement and to reveal patterns in such changes that could indicate flaws in training design or the lack of students' knowledge.

\subsection{Behavioral Clustering}

Clustering is essential to generic data mining. A clustering process discussed in~\cite{dutt2017} explains important steps in the design of clustering approaches in educational data mining (EDM). The recent survey paper~\cite{le2023} brings a useful classification of EDM tasks that use clustering models. Also, many specific approaches can be found in the literature addressing, for instance, the classification of students of small online courses by features adopted from business systems~\cite{wang2021} or revealing patterns of engagement in massive open online courses~\cite{khalil2017}.

In cybersecurity education, Svabensky et al.~\cite{svabensky2022} applied pattern mining and clustering techniques to analyze the usage of command-line tools in hands-on cybersecurity exercises, aiming to support the automated assessment of students. Burska et al.~\cite{Burska2024} used clustering directly to reveal gameplay strategies and possible flaws in the training content. Our approach utilizes process models to capture static patterns of behavior, i.e., typical or exceptional traces of activities. On the contrary, clustering is used to classify dynamic changes in behavior over time. Nevertheless, the analyst can combine these views, i.e., to pay attention to similar traces with similar dynamic changes.

\section{Properties of Process Graphs} \label{sec:rq1}

The size, complexity, and other factors can significantly influence the interactive techniques used to visualize process graphs~\cite{nobre2019}. This section analyzes the characteristics of event logs and their corresponding process models to establish fundamental design constraints for visualizing process graphs as multivariate networks.

\subsection{Quantitative Analysis}

To analyze these properties, we used publicly available datasets from two typical training events~\cite{CTFdata}. Both datasets contain three types of events. Game events track activities related to the training scenario and user progress, such as session start, hint usage, or successfully solved cybersecurity tasks. The other two event types -- Bash commands and Metasploit commands -- record trainee interactions with the network infrastructure. They capture specific commands and their parameters that were executed on the command lines of network nodes.

The first dataset included 52 trainees, while the second had 48 participants. In total, 11757 events were collected. Of these, 3597 were training progress events, 5669 were Bash commands, and 2491 were Metasploit commands.

\begin{figure*}[ht]
  \centering
  \includegraphics[width=1\textwidth]{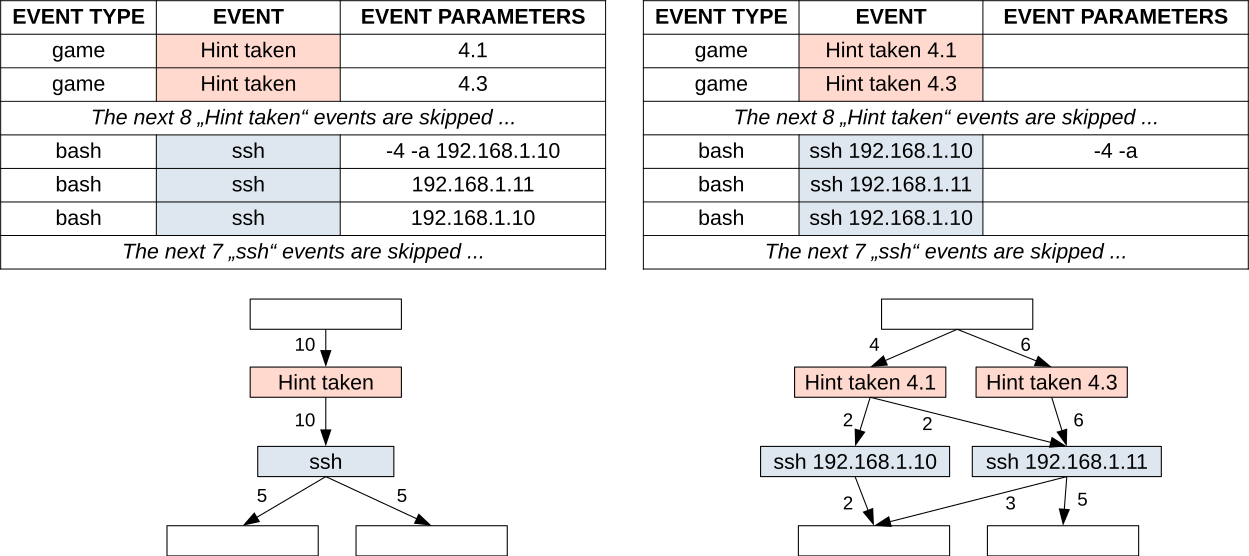}
  \caption{The impact of parsing and mapping event logs on graph complexity. Since only data from the \emph{EVENT} column is transformed into process mining activities and represented as graph nodes, the overall graph complexity depends on whether additional parameters from the raw event logs are also included in these activities.}
  \label{fig:heuristic-net-complexity}
\end{figure*}

The size of graphs generated through process discovery largely depends on how raw data is mapped onto process mining activities~\cite{PMinKYPO1,PMinKYPO2}. Bash and Metasploit commands introduce high variability, increasing the average affinity of the process.

For instance, if all executions of the \emph{ssh} command are treated as a single activity, regardless of additional arguments, the resulting graph contains only one node, as shown in the left-hand example in Figure~\ref{fig:heuristic-net-complexity}. Conversely, if \emph{ssh} commands with different remote addresses are considered distinct activities, multiple nodes are generated, each representing a unique \emph{ssh} connection, e.g., ``ssh 192.168.1.10'' and ``ssh 192.168.1.11''. Other optional parameters possibly used by users, e.g., ``-4 -a'' in the example, still do not affect the graph structure.

There is no universally correct or superior mapping approach, as the choice depends on the specific analytical objectives. To account for this variability, we analyzed the lower and upper bounds of graphs derived from the real datasets using different event mapping strategies. Table~\ref{tab:graph-size} summarizes the observed approximate minimum and maximum number of nodes in the resulting graphs.

%
%
\begin{table}[!htbp]
    \caption{Estimation of the size of process mining graphs.}
    \centering
    \begin{tabular}{|r|ccc|ccc|}
        \hline
        & \multicolumn{3}{c|}{\textbf{Dataset 1}} & \multicolumn{3}{c|}{\textbf{Dataset 2}} \\
        & \makecell{raw \\ events} & \makecell{upper \\ bound} & \makecell{lower \\ bound} & \makecell{raw \\ events} & \makecell{upper \\ bound} & \makecell{lower \\ bound} \\
        \hline
        Bash        & 2749 & 1322 & 207 & 2920 &  981 & 137 \\
        Metasploit  &  904 &  291 &  86 & 1587 &  350 & 108 \\
        Game events & 1617 &  146 & 146 & 1980 &  124 & 124 \\
        \hline
        \textbf{Total} & 5270 & \textbf{1759} & \textbf{439} & 6487 & \textbf{1455} & \textbf{369} \\
        \hline
    \end{tabular}
    \label{tab:graph-size}
\end{table}

The \emph{raw events} are introduced only for completeness. This column captures all events in the dataset, including noise and duplicates that have to be removed during preprocessing before performing any serious analysis~\cite{PMinKYPO2}. On the contrary, the \emph{upper bound} and \emph{lower bound} columns represent more realistic estimations of the number of nodes in process graphs. 

For the estimation of upper bounds, the commands, including their parameters, were mapped to process mining activities directly. For instance, \emph{ls~-a} and \emph{ls~-al} records are considered different Bash commands in this upper bound analysis, although they do almost the same. On the other hand, all dummy, duplicate, and noisy data were removed. In particular, we identified a lot of noise in Bash commands caused by copying data into the terminal. Apparently, some trainees accidentally copied the content of some files, e.g., public SSH key, into the terminal, causing each line to be recorded as an individual command with no semantic meaning for analysis. 

Like upper bounds, the lower bound numbers were also measured with any dummy, duplicate, and noise data removed. However, only the core commands without parameters, e.g., \emph{ls}, were mapped to process mining activities. In this case, the process graphs are significantly simpler, but a lot of details may be missing that could be important for analysts.

In contrast to Bash and Metasploit commands, the variability of game events is minimal. Therefore, the upper and lower bounds are the same for this type of event.

\subsection{Summary}

The analysis has shown that the real size of process graphs may significantly vary depending on which raw events are available and which of them are embraced in the analytical workflow. The upper and lower bounds indicate that the size fits medium ($< 1.000$) and large ($> 1.000$) size categories of Nobre's classification. The medium size can be achieved if the analysis focuses on certain aspects of the gameplay or when dealing with only selected event types. On the contrary, overview graphs capturing the whole game can reach thousands of nodes. Even worse, our experience has shown that drilling down into training details, e.g., exploring how a certain puzzle (cybersecurity task) was treated by trainees, small-size graphs ($< 100$) are often used for learning analytics. Therefore, analysts can meet all size categories when exploring training data. As there is no ideal solution, a visual-analysis tool should adapt to the size of process graphs.

\begin{table}[!htbp]
    \caption{Summary of process graph features.}
    \centering
    \begin{tabular}{|c|c|c|c|c|}
        \hline
        \textbf{size} & \textbf{type} & \textbf{\makecell{node \\ attributes}} & \textbf{\makecell{edge \\ attributes}} & \textbf{\makecell{topological \\ structure}} \\
        \hline
        \makecell{small \\ medium \\ large} & k-partite & heterogeneous & homogeneous & \makecell{path \\ single node/edge} \\
        \hline
    \end{tabular}
    \label{tab:graph-features}
\end{table}

With respect to this observation, other aspects of multivariate networks must be considered when designing the VA tool. They are summarized in Table~\ref{tab:graph-features}. The puzzle-based structure of training content, where each task is finished by a milestone (a flag), produces significant k-partite graphs, enabling us to split large graphs into middle or small-size sub-parts easily. Node attributes are heterogeneous because nodes encode varying trainees' activities. On the contrary, edge attributes are homogeneous, as they only encode the number of traces or unified time information. 

Regarding significant topological clusters, process graphs are strongly path-related as they reconstruct traces of behavior, whereas details of a single node or edge provide important insight into the behavior. We found other topological structures defined in~\cite{nobre2019}, i.e., communities (clusters of strongly connected nodes) and networks/sub-networks, less useful for the learning analytics of CTF process graphs. 
Networks/sub-networks are delimited by the strong k-partite structure of process graphs produced by puzzle-based training content. Clustering is important, but in a different context: analyzing the similarity of traces and then identifying typical or exceptional behavior.

\section{Analytical Objectives} \label{sec:rq2-objectives}

Based on the discussion with domain experts -- developers of hands-on cybersecurity exercises and organizers of training sessions, we formulated three generic objectives that can be used for the initial exploration of training data, aiming to identify any significant or suspected behavior or flaws in the training design.
\begin{enumerate}
    \item[O1] \emph{Distinguishing between typical and exceptional behavior.} Both aspects can reveal important features of training design or gameplay strategies. Outliers can indicate talented or cheating individuals, while the similar behavior of multiple trainees can be used to prove the clarity of training instructions, for instance. Depending on the purpose, identified groups or individuals can be further explored or, on the contrary, temporarily hidden from the analysis. Therefore, visual awareness of individual vs. collective behavior should be supported.
    \item[O2] \emph{Performance.} Supervised training is time-limited. However, estimating the time requirements in advance is very difficult. Moreover, it may differ even for the same training scenario if the skills of trainees are different. Therefore, the knowledge of how much time trainees have spent solving different tasks can help in analyzing the temporal aspects of training sessions.
    \item[O3] \emph{The change of behavior in time.} The behavior of trainees can change over time due to frustration, unclear instruction, lack of time, or lack of knowledge. It is necessary to capture such changes and trends to identify time points or spans when the change happened and potentially analyze the cause. Together with \emph{O1} and \emph{O2}, it would be possible to judge if it is the exceptional change or something more serious that affected the majority of trainees. Therefore, it is necessary to include time-evolving views and interactions in the VA tool.
\end{enumerate}

An analytical tool that addresses these objectives could effectively prove or disprove higher analytical hypotheses. In what follows, we discuss three examples of typical post-training questions that appear to be often asked by analysts of CTF games.

\paragraph{Engagement} Although the tasks of puzzle games are strictly defined, there is always a certain freedom in how to finish the task. Trainees can make an effort to find a solution by themselves or choose the simplest way by taking hints, for instance. This engagement can be affected in the very beginning (the trainee was forced to conduct the training without any personal interest), but also often changes during the training for many reasons, e.g., the lack of time or flaws in training design or infrastructure.
While the active usage of commands (either correct or incorrect) can indicate high engagement, taking hints or revealing a complete solution usually indicates a loss of motivation. Involving these types of events in the \emph{O1} and \emph{O3} and putting them into the temporal context \emph{O2} can help to prove or disprove hypotheses about engagement and reasons for its loss.
    
\paragraph{Cheating} Possible cheating can be recognized by super-performance \emph{O2} (the tasks are solved very quickly), exceptionally high assessment (if assessment data are present in the dataset), or exceptional behavior (unusual traces of activities) \emph{O1}. Partial cheating, i.e., cheating only at a certain part of the training or since a certain time, also needs the involvement of \emph{O3}. 
    
\paragraph{Difficulty} The estimation of the difficulty of tasks is crucial for assigning appropriate time and assessment rules so that trainees stay engaged in the training. The overall preview can be obtained from performance \emph{O2} in combination with behavioral clustering \emph{O1}, aiming to compare the behavior of the majority with outliers (either exceptionally skilled or weak individuals). The differences in the difficulty of individual tasks and then the changes in the difficulty during the gameplay can be observed from \emph{O3}.

\section{Dashboard Design} \label{sec:rq3-solution}

\begin{figure*}[ht]
  \centering
  \includegraphics[width=\textwidth]{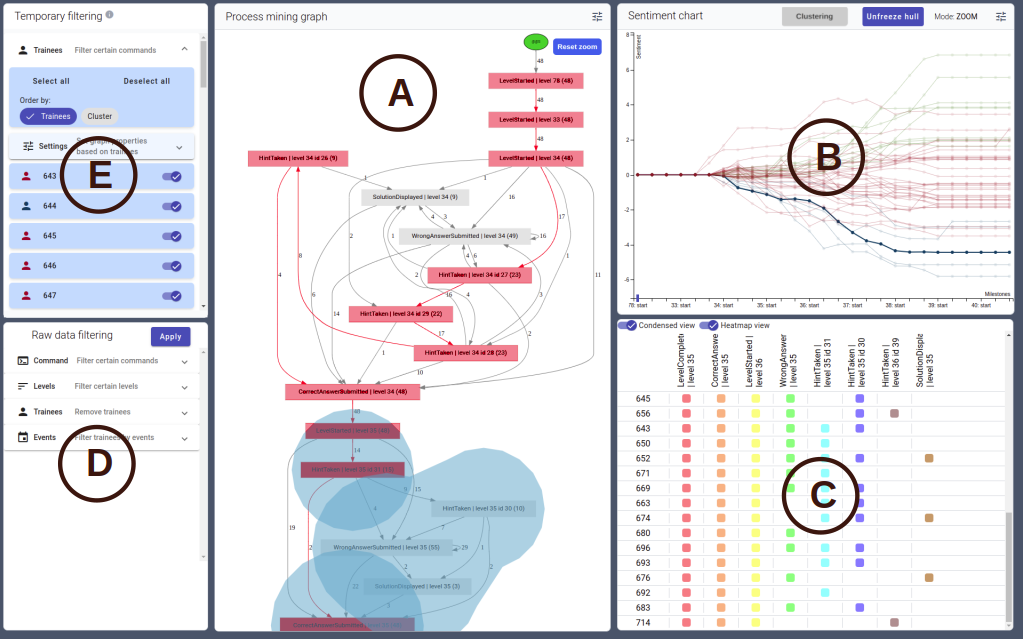}
  \caption{Overview of the analytical dashboard: Process mining graph (A), sentiment views (B), details of individual walkthroughs (C), raw data filtering (D), and temporary visual suppression (E).}
  \label{fig:dashboard}
\end{figure*}

This section summarizes the visual analytics dashboard proposed to support exploratory tactics addressing objectives \emph{O1 -- O3} and reflecting design constraints derived from the analysis of process graph properties. The layout of the dashboard depicted in Figure~\ref{fig:dashboard} is built around the obvious graph representation of process models (A) complemented by visual tools for temporal behavioral analysis, aka sentiment views (B), tools for individual walkthroughs (C), data filtering (D), and visual suppression (E). In what follows, we discuss the functionality of individual parts, as well as their interconnection and design decisions.

\subsection{Process Graphs}

\begin{figure*}[ht]
  \centering
  \includegraphics[width=0.8\textwidth]{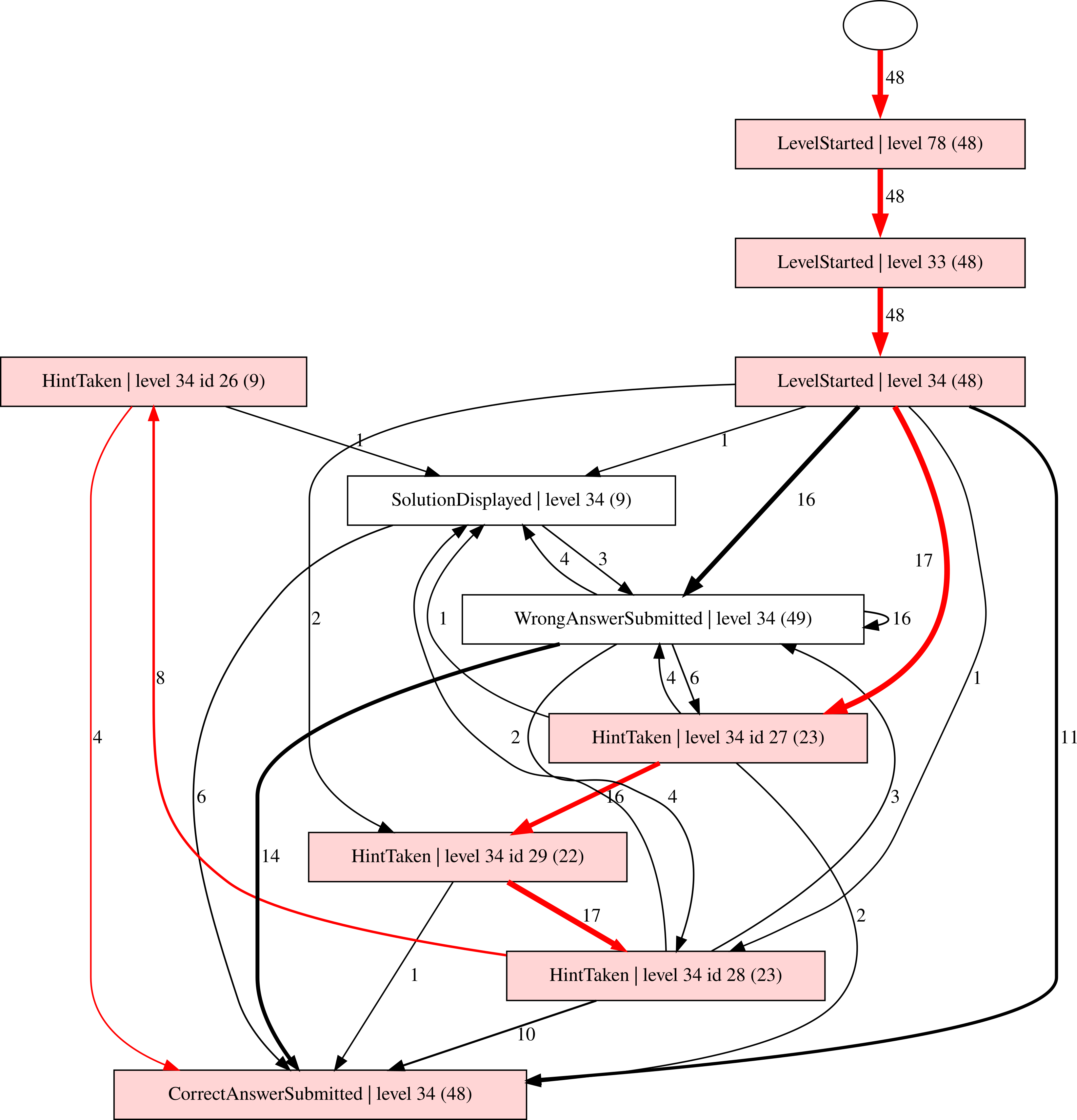}
  \caption{Frequency graph highlighting the importance of traces by bold arrows and activities of selected trainees by color-coding (red arrows and nodes in the graph).}
  \label{fig:heuristic-net}
\end{figure*}

Due to the process-oriented nature of practical cybersecurity exercises, the prominent part of the analytical dashboard is occupied by the process graphs (Figure~\ref{fig:dashboard} (A)) constructed from event logs automatically using obvious methods of process discovery. It has been shown that heuristic nets fit the analytical goals of CTF educators the best~\cite{PMinKYPO1}. Obtained frequency graphs inform the analyst of the number of trainees following the same path, as shown in Figure~\ref{fig:heuristic-net}. 

The plain visualization of the oriented process graph is often decorated by various visual effects, aiming to attract the analyst's attention to significant parts of the graph. We combine several complementary visual principles. 

The thickness of the edges corresponds to the number of traces. Moreover, traces (nodes and edges) of a particular trainee can be highlighted on the graph by using different color coding so that the analysts can identify specific activities of a selected individual. In Figure~\ref{fig:heuristic-net}, the path of the selected trainee is rendered in red. Together, this kind of visual highlighting enables the analysts to glimpse typical and rare walkthroughs and then address the objective \emph{O1}). 

\begin{figure*}[ht]
  \centering
  \includegraphics[width=0.8\textwidth]{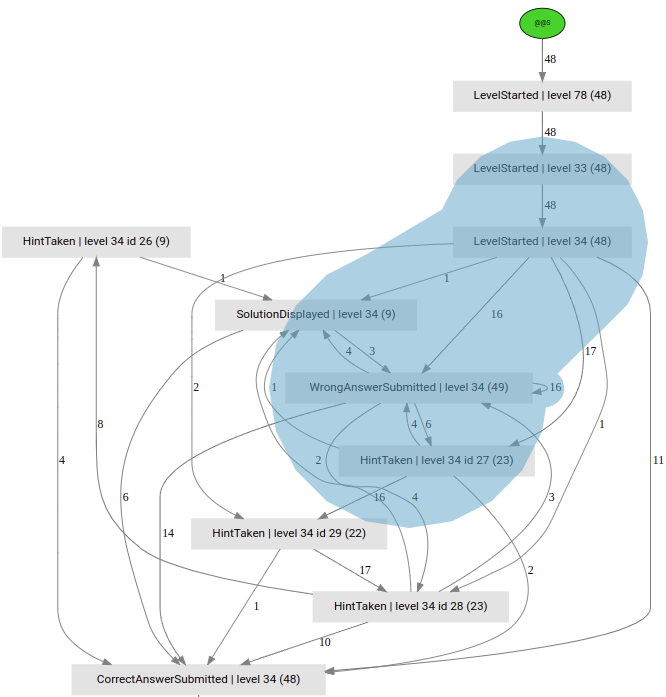}
  \caption{Frequency graph highlighting activities conducted by multiple trainees at similar times.}
  \label{fig:heuristic-net-hulls}
\end{figure*}

However, the path highlighted by their importance or relevance to selected trainees does not consider possible time constraints (the objective \emph{O2}). For performance analysis, the analyst needs to identify whether some activity was performed sooner or later by one trainee than another. Or whether it was performed at the same time. Therefore, we introduced the so-called \emph{temporal proximity of activities}, where events that appear in a user-defined time span are considered activities performed at the same time. Given a time point, we can highlight such nearby activities on the process graph. Figure~\ref{fig:heuristic-net-hulls} demonstrates the visual encoding. Activities included in the blue area were performed ``around the same time'' selected by the analyst. We use convex hulls computed around the corresponding nodes to avoid visual interference with other highlighting approaches described above. This is feasible because nodes representing activities close in time are typically positioned close to each other in the process graph and then localized in a specific part of the graph. 

\begin{figure*}[ht]
  \centering
  \includegraphics[width=0.5\textwidth]{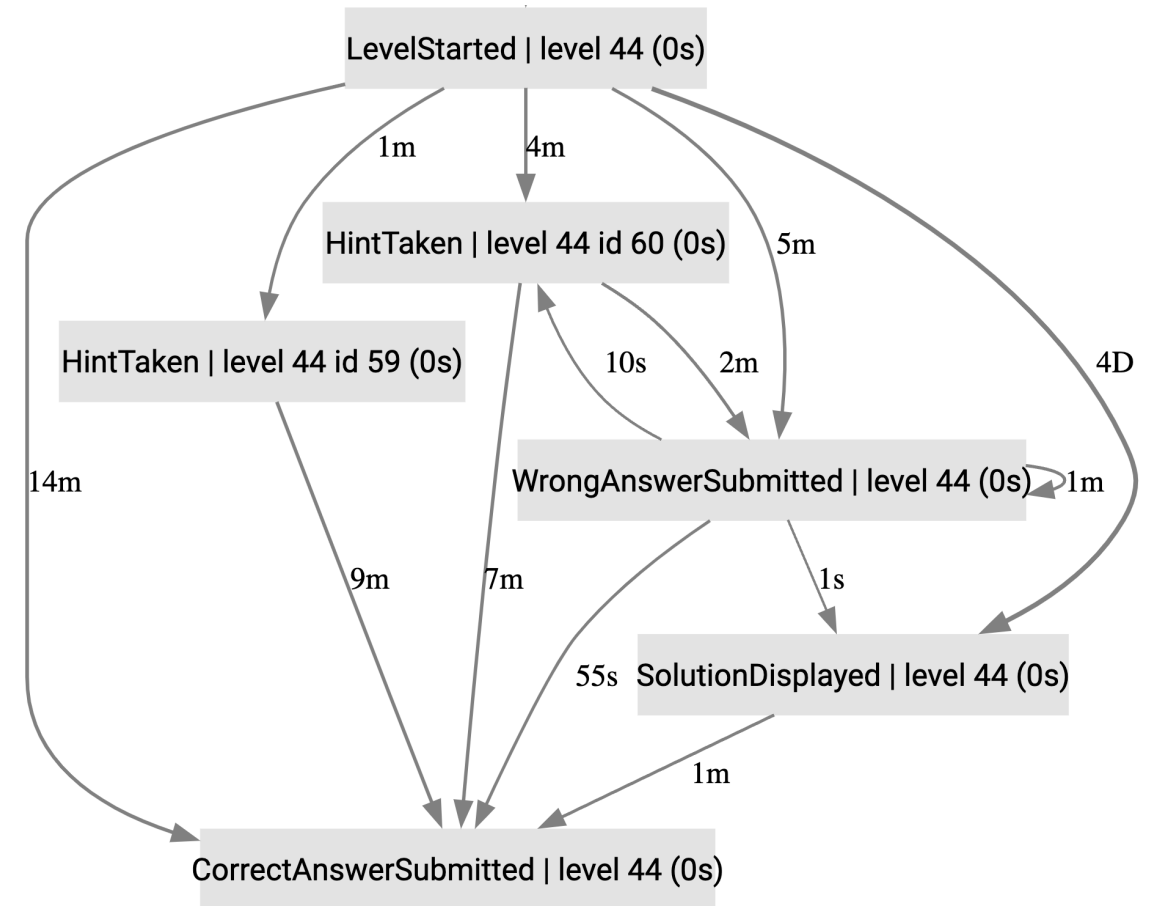}
  \caption{Performance graph, where numbers on edges encode time spent by activities.}
  \label{fig:performance-graph}
\end{figure*}

Another approach to exploring time aspects of the training (and then addressing the objective \emph{O2}) is based on switching the frequency-based heuristic net to the so-called performance graph that replaces the number of trainees on edges with the average, median, maximal, or minimal time required by trainees to proceed between nodes, as shown in Figure~\ref{fig:performance-graph}. This view can be used to distinguish between simple and time-demanding tasks of the exercise. 

It appeared that there was no need to show performance and frequency variants of process graphs simultaneously. The decorated performance graphs are suitable mainly for the exploration of trainees' walkthroughs, while frequency graphs are more appropriate for identifying time bottlenecks in training scenarios. Therefore, the graph view can toggle between the two variants, never displaying them side by side, saving space in the dashboard. Moreover, the view is intentionally located in the prominent middle part of the dashboard with vertical alignment that fits the elongated shape of many graphs given by the k-partite structure of the data. 

Despite the wide area allocated for process graphs, their possible complexity still remains an issue. Our experiments proved satisfying usability for restricted data views, e.g., the analysis limited to only a specific exercise level (puzzle) or the analysis of the whole exercise but restricted to selected event types only. Otherwise, the obtained process graphs may become too complex and then incomprehensible.
Therefore, we proposed two strategies to tackle generic graph complexity. First, the analyst should be able to restrict the data interactively during the analysis. The data filtering and temporary hiding are discussed below. Second, the dashboard could provide different views for different levels of abstraction and enable the analysis to drill down into details only if necessary. The idea lies in replacing a detailed k-partite graph with a simplified overview, where sub-graphs are replaced with a series of statistical previews.

\begin{figure*}[ht]
  \centering
  \includegraphics[width=\textwidth]{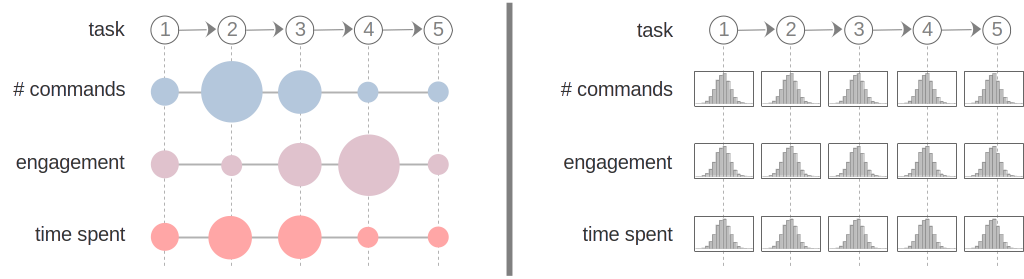}
  \caption{Examples of a statistical overview of process graphs, where sub-graphs are replaced with statistical values. Either simplified (left) or more detailed (right) visual encoding can be used.}
  \label{fig:graph-overview}
\end{figure*}

Figure~\ref{fig:graph-overview} shows an example where five consequent gaming puzzles (i.e., cybersecurity tasks) that normally form sub-graphs of the complete process model are replaced with aggregated statistical views, e.g., the average number of used commands. The visual encoding may differ depending on analyst requirements. The left view simply uses the size of circles to visually encode the measured values, while the right-hand side example provides a more detailed statistical distribution of values, e.g., histograms. As these visualizations are very compact, they can include multiple metrics, e.g., the average number of commands used, trainees' engagement, or the relative time spent by trainees solving the task. Regardless of visual encoding, the visualization can be adaptive, enabling the analyst to show different metrics (rows) and drill down into detailed process graphs.

\subsection{Sentiment View}

Sentiment view is a term used in this paper for analytical visualizations dealing with dynamic changes in trainees' behavior, possibly reflecting changes in mood, sentiment, or other types of mental state  (Figure~\ref{fig:dashboard}(B)). Its primary purpose is twofold: Visually detecting changes in behavior over time and classifying trainees into groups based on these changes.

\subsubsection*{Sentiment Metric}

Frequency process graphs can reveal behavioral patterns. However, they provide only a static view aggregated from the behavior of multiple trainees. To address the objective \emph{O3}, we need a compact visualization emphasizing changes in the behavior of individual participants in time. Therefore, we proposed an algorithm that analyzes changes in trainees' behavior in time and transforms them into a visual representation.

The algorithm is based on a user-defined sliding time window in which a behavioral sentiment metric is computed. The window sequentially scans the entire span of the exercise in user-defined steps. The changes in the metric values become the subject of analytical visualizations.

The window size and the step size are adjustable. They are expressed as percentages of the measured lengths of game levels (puzzles), adapting the results to variable time requirements of cybersecurity tasks.

For the $i$-th time window, the metric value is computed as follows. First, a score is computed for each trainee as the sum of the number of weighted occurrences of involved events:
\begin{equation} \label{eq:score}
    score_i = \sum_{e \in events} w_e \cdot n_{e,i}, 
\end{equation}
where $w_e$ detonates the weight of the event $e$ and $n_{e,i}$ the number of occurrences of $e$ in the time window $i$. Then, the values are normalized to the range from $-1$ to $1$ using linear interpolation to the following range:
\begin{equation} \label{eq:normalization}
    median(score_i) \mp max(score_i),
\end{equation}    
where $median$ and $max$ denote median and maximal score values among all trainees.
Using this metric, a negative value represents lower performance by the trainee (e.g., lower engagement) compared to other trainees at that moment, and vice versa.

Drawing the values cumulatively provides insight into the changes in the measured trainees' behavior in time, as shown in Figure~\ref{fig:dashboard}(B) and in detail in Figure~\ref{fig:case-study}. The X-axis represents a timeline that includes steps of the sliding window. The Y-axis includes the cumulative values of the metric measured in the sliding window for each trainee. Ascending trends indicate an increase in the measured performance. Measured values are depicted as points on the lines. However, because the graph is dense, the close points are replaced with a single joint point, reducing visual complexity. 

Sentiment view also serves as the selection tool for process graphs. When the mouse hovers over points of the sentiment chart, corresponding trainees and activities close in time are selected and highlighted on the graph view. Moreover, the graph is zoomed into the convex hull area, providing a detailed view of possible complex graphs. Moving the mouse over the timeline of the sentiment chart can produce an animation of convex hulls, enabling the analyst to ``replay'' trainees' activities when their sentiment changes.

\subsubsection*{Clustering} 

Generic clustering methods have been shown to support visual learning analytics~\cite{Burska2024} in practical cybersecurity education in general. We use them to cluster dynamic changes in behavior (the objective \emph{O1}), i.e., the values computed by behavioral metrics and presented on the cumulative graph. Two different clustering views are proposed. 

\begin{figure*}[ht]
  \centering
  \includegraphics[width=1\textwidth]{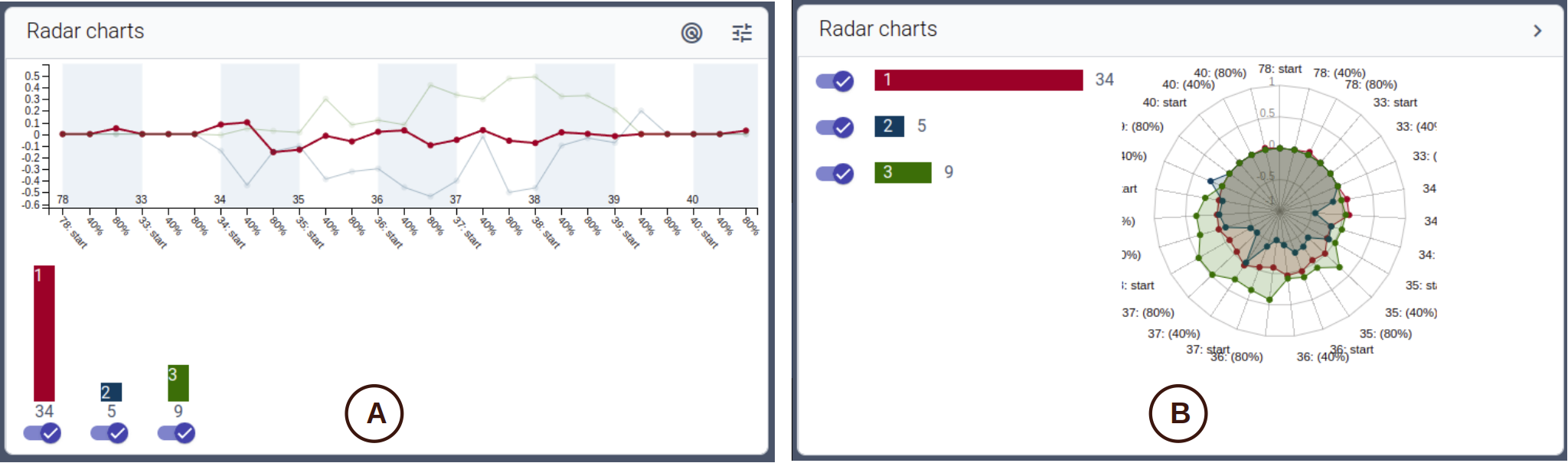}
  \caption{Clustering trainees by their changes in behavior. Line chart (A) and spider chart views (B).}
  \label{fig:graph-clustering}
\end{figure*}

The line chart view depicted in Figure~\ref{fig:graph-clustering} (left) is visually and functionally identical to the cumulative graph of the individual walkthroughs. It is more sparse as now the lines group individual trainees with similar shapes, indicating similar changes in behavior. A simple bar chart shows how many trainees belong to clusters. Whole clusters (their trainees) can be filtered out from the analysis using switches under bars.

The second view depicted on the right-hand side in Figure~\ref{fig:graph-clustering} has the form of an obvious spider chart. This visualization better shows relationships between clustering features, possibly revealing patterns in changes in the behavior. 

Clustering visualizations are not visible in the dashboard overview in Figure~\ref{fig:dashboard} because they occupy the sector (B) together with the previous visualization. This is because they are never used simultaneously. They represent different tactics to analyze and filter dynamic changes in trainees' individual or collective behavior. Therefore, these views can be switched easily, and only one view is visible at the time.

Also, broad configuration options are available, including the elbow function analysis to estimate an appropriate number of clusters. Currently, we use an unsupervised k-means method as the clustering algorithm. Although the results have shown good results from the perspective of generated clusters~\cite{Chudovsky2023thesis}, the k-means approach suffers from randomness. It can produce different clusters for every invocation and then confuse the analyst. Therefore, other methods, e.g., tree-based clustering, could be used instead. On the other hand, these methods are often supervised, which can pose problems for varying exercises with only a few realized training sessions.

\subsection{Detailed Traces of Activities}

Highlighting the paths on the process graphs enables analysts to pay attention to selected trainees and their activities. Still, the data are aggregated. The view shows how many trainees used the given command (in the case of the frequency graphs) or how long they spent on the activity (in the case of the performance graphs). However, for an in-depth investigation, an analyst has to be able to distinguish between the activities of individual trainees.

We introduced a compact view that complements the process graphs by decomposing individual paths into per-trainee details. The visualization shown in Figure~\ref{fig:dashboard}(C) has the form of a matrix where rows represent individual trainees, columns are data events, and cells provide necessary details. Currently, simple color dots are used to indicate the presence or absence of per-trainee activities. Details, e.g., complete commands with parameters, are provided as tooltips. This view helps tackle the performance of individuals (the objective \emph{O2}) and compare them with others (the objective \emph{O1}).

The matrix is fully connected with other views, so it always displays relevant data depending on constraints put on trainees, activities, and time spans triggered by other dashboard parts. For example, suppose the analyst selects trainees and a time span in the sentiment view. In that case, relevant paths are highlighted on the process graph, the process graph is zoomed into the corresponding activities (the blue convex hulls), and the matrix shows commands of related trainees. In this way, the analyst can dynamically explore trainees' behavior just by moving the mouse pointer over the sentiment chart and observing all relevant data side-by-side. On the other hand, the interaction can be frozen for a while in the sentiment view so that the analyst can fix the views and conduct an in-depth exploration, e.g., moving and zooming the process graph or exploring the matrix of trainees' commands.

\subsection{Raw Data Filtering}

The data analysis discussed in Section~\ref{sec:rq1} has shown that events captured during the real training sessions can include meaningless records or data unrelated to analytical objectives. Therefore, we proposed an interactive filtering user interface that enables analysts to tackle information and visualization complexity already at the level of raw data filtering. The interface, shown in Figure~\ref{fig:dashboard}(D), reflects multiple data abstractions observed during the analysis of event types and their characteristics.

\begin{figure*}[ht]
  \centering
  \includegraphics[width=1\textwidth]{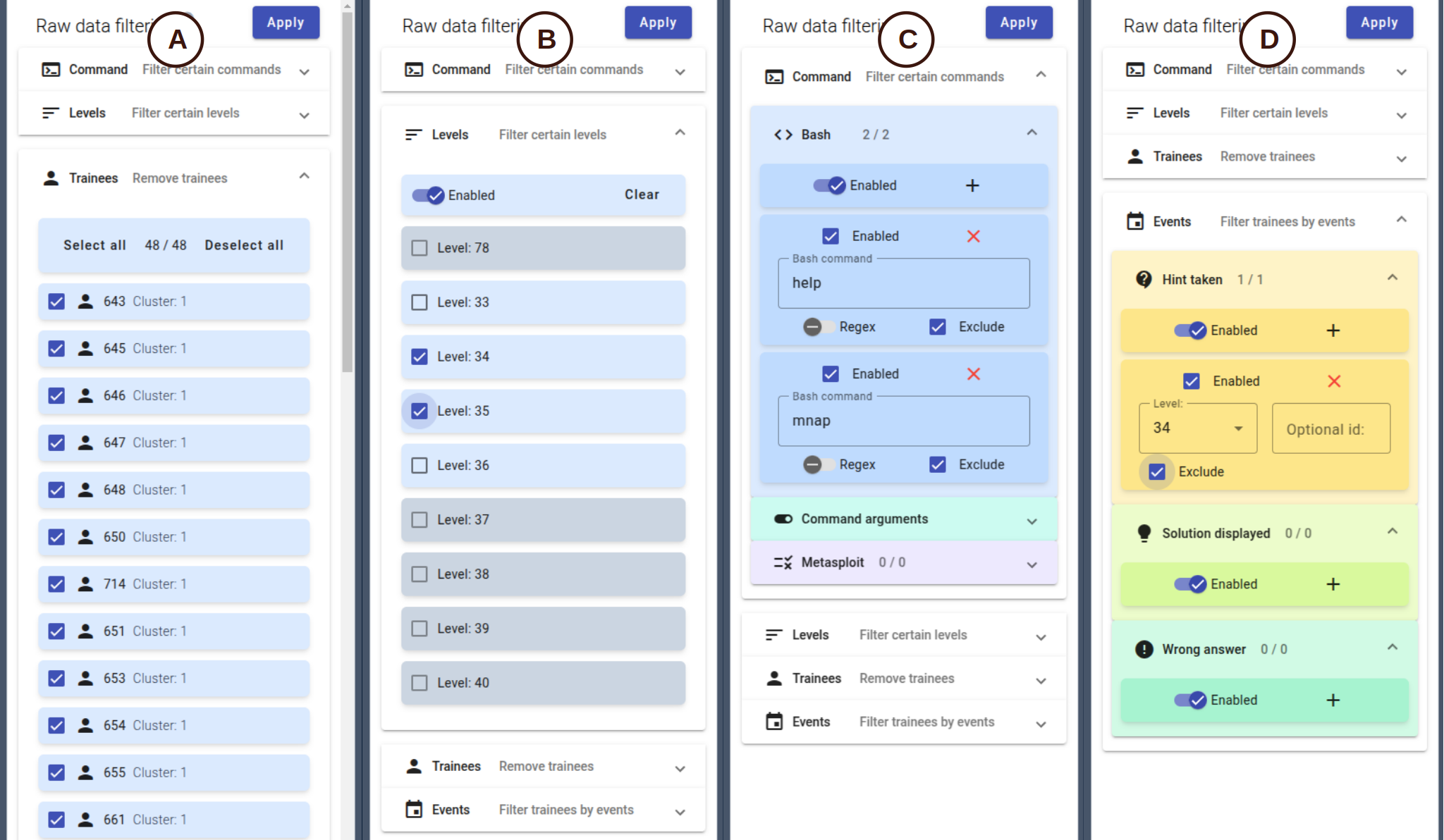}
  \caption{GUI for raw data filtering. Examples for trainees (A), training levels (B), commands (C), and game events (D).}
  \label{fig:raw-data-filtering}
\end{figure*}

\paragraph{Filtering by trainees} Implemented as a simple select box that enables the analyst to exclude data of particular trainees from the analysis, as shown in Figure~\ref{fig:raw-data-filtering}(A). It can typically happen if the collected data contains some testing walkthroughs from dry runs that should be ignored. All trainees are included by default. 

\paragraph{Filtering by training levels} Implemented also as a simple select box (Figure~\ref{fig:raw-data-filtering}(B)). Data from unchecked game levels (training tasks) are removed from the analysis. However, we found it very confusing if non-subsequent levels were chosen, omitting some levels between them. Therefore, the selection of only subsequent levels is possible, e.g., levels 2 \& 3 \& 4, but not 2 \& 4.

\paragraph{Filtering commands} The data analysis revealed the presence of many noisy events that are usually caused by copying a file or web content to the terminal. The content is then recorded as many Bash or Metasploit garbage events. Therefore, the filtering tool supports composing boolean expressions that exclude Bash or Metasploit events with given content from the analysis (Figure~\ref{fig:raw-data-filtering}(C)). Multiple rules can be added and removed dynamically. They are interpreted as logical \emph{AND} between them. Each rule defines a text representing either a keyword or a regular expression. Events containing the text are either included in or excluded from the analysis. 
Besides filtering the whole Bash and Metasploit commands, the tool also supports user-defined mapping of commands and their arguments onto process mining activities, as discussed in Section~\ref{sec:rq1}. Therefore, it is possible to interactively define rules that distinguish between ``ssh root@172.18.1.5'' and ``kali@172.18.1.5'' commands while considering all other ssh commands the same activity, for instance.
Often-used rules could be predefined in the future. 

\paragraph{Filtering game events} While commands represent steps in solving cybersecurity tasks, game events can be rather perceived as the state of the gameplay, i.e., the progress in the training. Moreover, they are much more reliable without noisy data. This is why they are treated differently during the analysis. First, some game events can be considered mandatory because they are used to identify the progress of the trainee, e.g., the successful solving of a task. Filtering such events makes no sense. On the contrary, some expressions related to game events could be valuable, e.g., when focusing only on trainees who used hints at certain levels. Therefore, the filtering tool supports the filtering of events such as \emph{hints taken}, \emph{solution displayed}, and \emph{wrong answer} in specified game levels (Figure~\ref{fig:raw-data-filtering}(D)). The design of interaction follows the principles of command filtering, i.e., dynamically managed rules connected with implicit logical \emph{AND}.

\subsection{Temporary Visual Suppression}

While raw data filtering aims to completely exclude irrelevant data from the analysis, temporary filtering is intended to be used during the analysis to simplify the view of the data. It aims to attract analysts' attention to only relevant pieces of information while temporarily hiding or visually suppressing less important data. Nevertheless, temporarily suppressed data are still used in the underlying learning analytics, e.g., clustering or the computation of changes in behavior.

Showing only selected trainees has been identified as a very important requirement of analysts. Therefore, an easily accessible preview of trainees is exposed as a prominent part of the dashboard in the \emph{Temporary Filtering} panel (E) in Figure~\ref{fig:dashboard}. It has the form of a list sorted either by trainee names (IDs) or by behavioral clusters. The latter allows trainees with similar behavior to be shown or hidden quickly. 
The panel can be extended with other data types in the future if needed, following the already discussed raw data filtering design principles. 

\begin{figure*}[ht]
  \centering
  \includegraphics[width=1\textwidth]{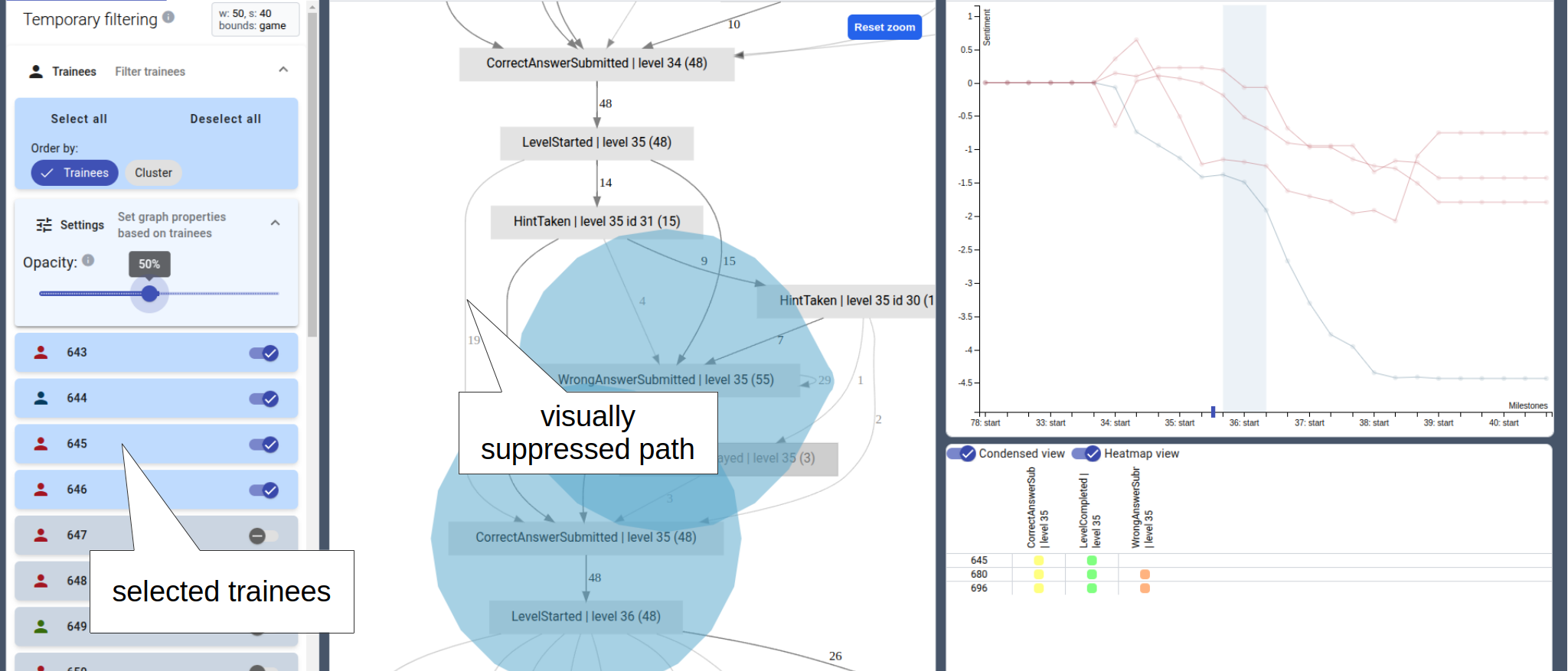}
  \caption{Temporary visual suppression.}
  \label{fig:visual-suppression}
\end{figure*}

The suppressed data are still shown in the analytical visualizations as they provide important context for analysts. However, they are shown less noticeably. Figure~\ref{fig:visual-suppression} depicts the situation when only four trainees are selected. Paths and activities of other trainees are still preserved in the process graph but rendered transparently. The strength of suppression (transparency) is maintainable. Behavioral clusters and sentiment lines are also computed for all trainees in the sentiment view. However, only selected trainees are shown.

\section{Case Study: Engagement Analysis} \label{sec:case-study}

A prototype implementation of the analytical dashboard is available as an open-source project developed at Masaryk University~\footnote{\url{https://gitlab.fi.muni.cz/kypo-analyst}}. 
This case study aims to demonstrate the ability of the dashboard to support training designers and supervisors of cybersecurity CTF exercises in analyzing post-training data by interactively combining and evaluating analytical objectives \emph{O1--O3}, as defined in Section~\ref{sec:rq2-objectives}. We chose engagement to showcase highly relevant hypotheses formulated at a higher level of abstraction.


Interactions and analytical workflows presented in this case study are backed by the initial qualitative evaluation that has been conducted to assess the overall usability of the dashboard and gain feedback from domain experts for further development. We involved five users who were experienced in supervising practical cybersecurity exercises organized in the KYPO Cyber Range Platform~\cite{Vykopal2017} or in CTF game design. Initially, we gathered basic information about the participants and their skills. Then, they familiarized themselves with the analytical tool. They were provided with an instructional video that explained the basic concepts and different types of visualizations, as well as guidance on configuration and filtering options. Finally, the participants were asked to complete four analytical tasks and answer questionnaires related to their findings, the difficulty of the tasks, and the usefulness of visualizations. The evaluation tasks are briefly summarized below, while the detailed information and supplementary materials can be found in~\cite{kadavy2025,maly2025}.
\begin{enumerate}
    \item[T1] Identify trainees with consistently low or high engagement throughout the exercise.
    \item[T2] Identify trainees whose engagement increased or decreased during the exercise.
    \item[T3] Verify hypotheses that a selected trainee's low engagement was due to the lack of time. 
    \item[T4] Verify hypotheses that a selected trainee's low engagement resulted from insufficient prior knowledge.
\end{enumerate}

\subsection{Dataset and Training Scenario}

The data used in this case study are publicly available as \emph{dataset1} from~\cite{CTFdata}. The related training scenario consists of the following steps:
\begin{itemize}
    \item After the introductory information and instructions, the trainees were given credentials for an attacker KALI Linux machine. The trainees were expected to use a tool \emph{nmap} with an option \emph{-sV} to scan the server and find the name of a vulnerable service running on an open port, which also served as a flag for this level. 
    \item In the next task, trainees were supposed to find a CVE code of the vulnerability to exploit. To find the code, trainees could search the internet or use other tools such as Metasploit -- a command-line cybersecurity tool often used to gain access to or information about vulnerabilities of potential targets. 
    \item The following task consisted of teaching the trainees how to use Metasploit and search for and apply exploits. The goal was to exploit the vulnerability CVE-2019-15107 that they had found previously and access the server. The flag was a five-digit number located in a file in the \emph{/root/} directory. Trainees found an exploit based on the name of the service, chose the correct one, and learned how to set up the options for the exploit, such as setting RHOSTS to the victim's IP address, before running it. 
    \item Trainees learned that the hacker group deleted all their files on this server. It is, however, probable that they backed up all their files to another server. The trainees' task was to explore the local $.bash\_history$ file and find a clue that could lead to the new IP address, mainly looking for the usage of the \emph{scp} command.
    \item The following task consisted of searching the currently exploited server for information, allowing them to access the new server. The task was to obtain the private SSH key on the server and crack its passphrase. Trainees were assumed to use a password cracking tool \emph{john} for a dictionary attack using a provided script and wordlist on the attacker's machine.
    \item The final level informed students that they connected to the other server and found a note from the hacker group. 
\end{itemize}

We used raw data for the evaluation, preprocessed according to the lower bound complexity estimation discussed in Section~\ref{sec:rq1}. It means that dummy, duplicate, and noisy data records were removed. On the contrary, commands were mapped to activities without parameters, i.e., the ``ssh 192.168.1.10'' and ``ssh 192.168.1.11'', for instance, were considered the same command by default, appearing in the process graph as a single node. Although the respondents were informed about the possibility of changing the mapping, they did not do that as it was not necessary for engagement analysis.

\subsection{Initial Settings of the Sentiment Metric}

Engagement can be considered sentiment, whose changes can be estimated from correct answers, wrong answers, taking hints, displaying solutions, and the active usage of Bash or Metasploit commands. A higher frequency of taking hints can indicate the trainees' loss of motivation to find a solution by themselves. Therefore, the default weight of this event was set to -5. Reading step-by-step solution instructions by the trainee can be considered a complete resignation to making an effort to solve the cybersecurity task. Therefore, this event type had the weight set to -20. Similarly, the weight of wrong answers was set to -1 as their frequent appearance in the time window can indicate frustration of the trainee, causing an engagement decrease. On the contrary, a high frequency of commands can be considered positive engagement with the weight set to 1 by default. For a typical one-hour CTF game, the optimal window size seems to be 50\% of the game level length and the step size 40\%. 

The metric values were determined only using several simple experiments. Further research is required to establish strong foundations for the system of meaningful metrics used in the accumulative behavioral graph. Although the weights are configurable in the GUI and then fully customizable, they were not changed by participants during the evaluation, as fine-tuning the engagement metric was not the objective of the testing.

It is to be pointed out that these metric settings aim to demonstrate the principles of capturing changes in trainees' engagement. Another metric combining different events with different weights can be used to reflect other moods.

\subsection{Detection of Low Engagement}

The identification of trainees with low engagement was covered by two evaluation tasks \emph{T1} and \emph{T2}. First, the respondents were required to identify trainees with low and high engagement throughout the entire game. This involved comparing individual performance (objective \emph{O2}) against that of other trainees (objective \emph{O1}). Second, respondents had to identify trainees who either lost or gained engagement during the training. This task focused on analyzing temporal aspects (objective \emph{O3}) in relation to individual performance (objective \emph{O2}). In both cases, the sentiment view served as a key entry point for process analysis, allowing users to explicitly track engagement changes, cluster trainees based on engagement levels, and explore detailed insights through other juxtaposed views on the dashboard.

\begin{figure*}[ht]
  \centering
  \includegraphics[width=1\textwidth]{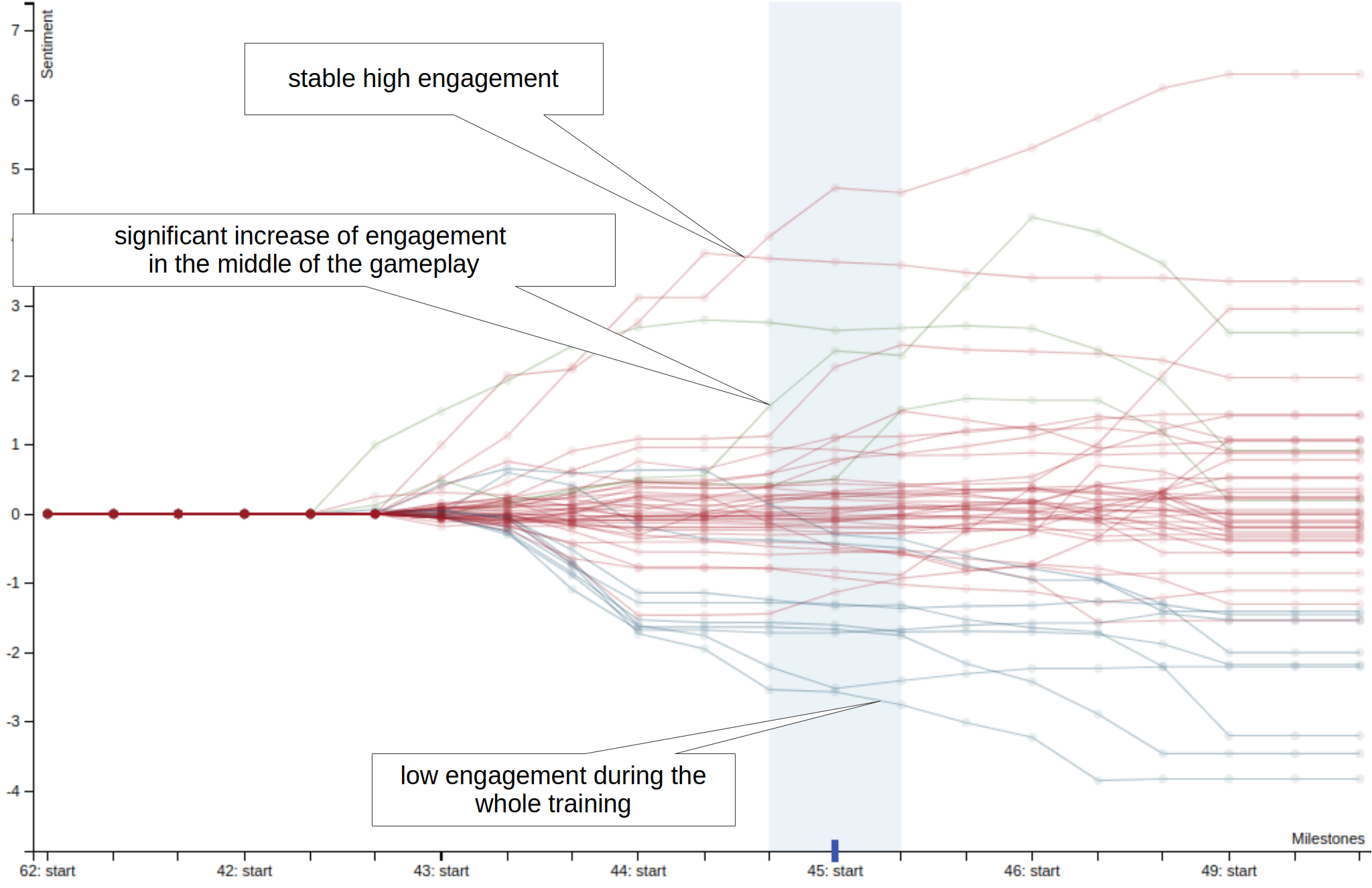}
  \caption{Sentiment view with marked examples of three types of engagement changes. The horizontal stripe suggests the time window delimiting nearby activities highlighted in other views.}
  \label{fig:case-study}
\end{figure*}

Most respondents followed this workflow, dominantly using the sentiment view with preset engagement metrics for their initial interaction. Three clusters were used by default, but the analyst could explore the elbow function and try different numbers of desired clusters. 

Figure~\ref{fig:case-study} shows the initial state of the sentiment chart. A quick inspection of the view reveals that several trainees from the same cluster experienced low engagement during the whole exercise. Two students maintained stable, high engagement during the exercise. For one student, the graph shows a significant increase in the middle of the exercise. Of course, the behavior of other trainees can also attract an analyst's attention. 

All respondents identified the same trainees with low and high engagement during the whole game. However, their answers diverged when it came to the second or third trainees with such characteristics. Moreover, the respondents disagreed in selecting trainees who either lost or gained engagement during the training. This inconsistency may originate from the multiple correct options, where the specific selection depends on the personal preferences of the analysts. Moreover, some participants interpreted the task as the selection of a single trainee, while others listed all trainees who exhibited changes in engagement.

Overall, respondents rated finding trainees with low, high, or changing engagement as an easy task and confirmed the usefulness of the sentiment chart for getting an overview of engagement in the training session.

\subsection{Inferring the Cause}

To investigate the possible reasons for the low, high, or changed engagement, the analysts have to combine multiple process views and interactions. 
In what follows, we turn our attention to the low-engaged trainee. The drill-down exploration of the cause typically starts on the sentiment chart, which appeared to be the favorite way to take an overview of trainees' behavior. The analyst can move the mouse pointer over the sentiment lines. Traces of activities are dynamically highlighted in the process graph. Simultaneously, corresponding events are updated accordingly in the detailed matrix view. This animation-like interaction helps the analyst distinguish between typical and exceptional behavior (objective \emph{O1}). For example, the analyst might observe that the blue areas in the process graph usually remain compact throughout the interaction. This suggests that the corresponding trainees exhibit similar behavior, as their events were recorded at similar times. In contrast, moments where these areas spread across distant parts of the graph indicate those parts of the training session where the behavior of selected trainees diverged. Moreover, data of other participants can be suppressed to reduce the visual complexity of visualizations, e.g., restricting the views to specific game levels (cybersecurity tasks) only.

\begin{figure*}[ht]
  \centering
  \includegraphics[width=1.0\textwidth]{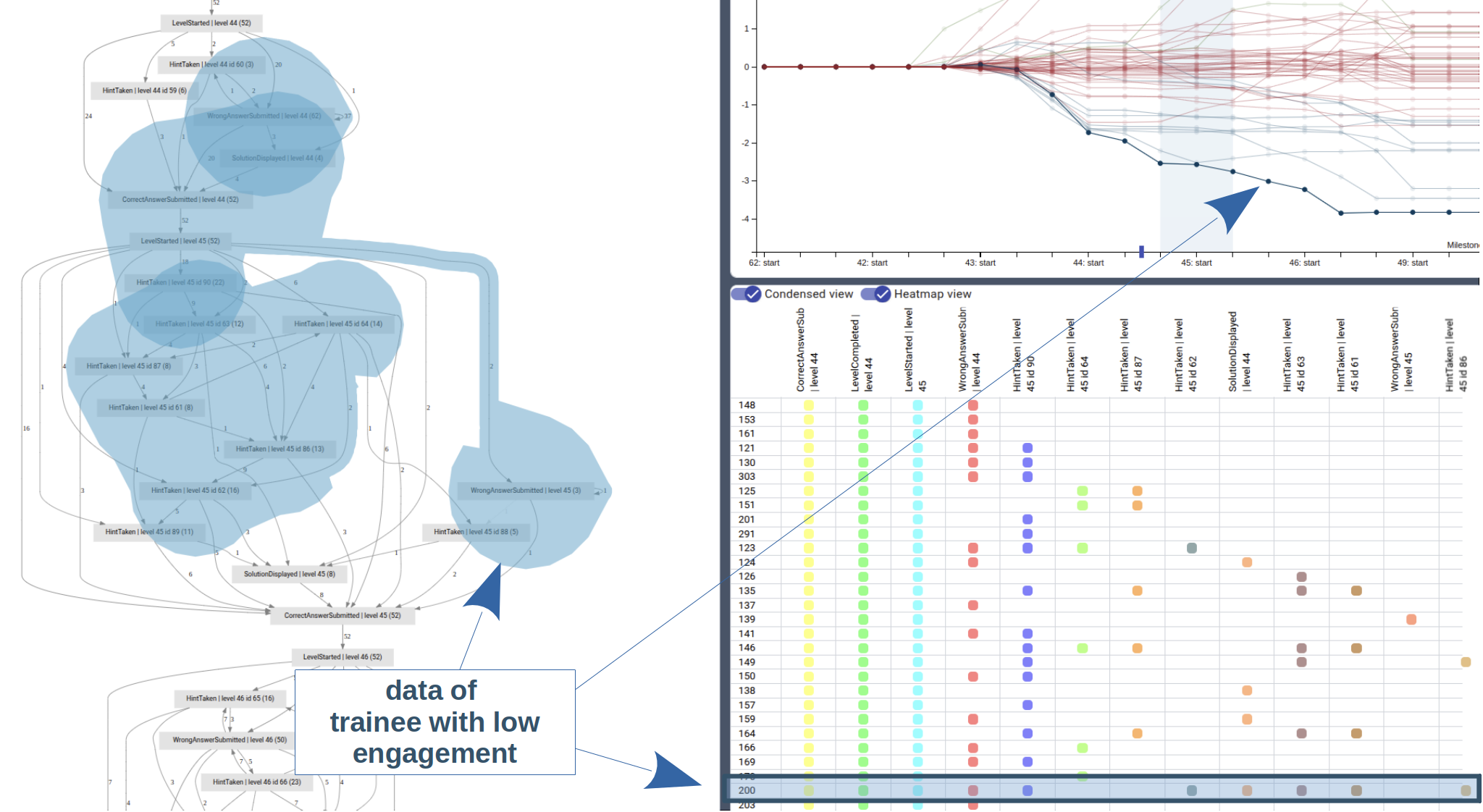}
  \caption{Behavioral data of the trainee with low engagement spread across juxtaposed views.}
  \label{fig:cause-analysis}
\end{figure*}

Combining animation-like interaction with the sentiment chart and information suppression can help analysts efficiently investigate specific trainees in detail (objective \emph{O3}). Figure~\ref{fig:cause-analysis} marks data of the trainee with low engagement spread across juxtaposed views. The event matrix in the bottom-right corner reveals that, compared to other trainees, this individual frequently relied on hints and often displayed complete solutions. Although Bash and Metasploit commands are not visible in the screenshot, enabling them in the visualizations would reveal that, with few exceptions, the trainee did not actively execute any meaningful commands.

\begin{figure*}[ht]
  \centering
  \includegraphics[width=0.5\textwidth]{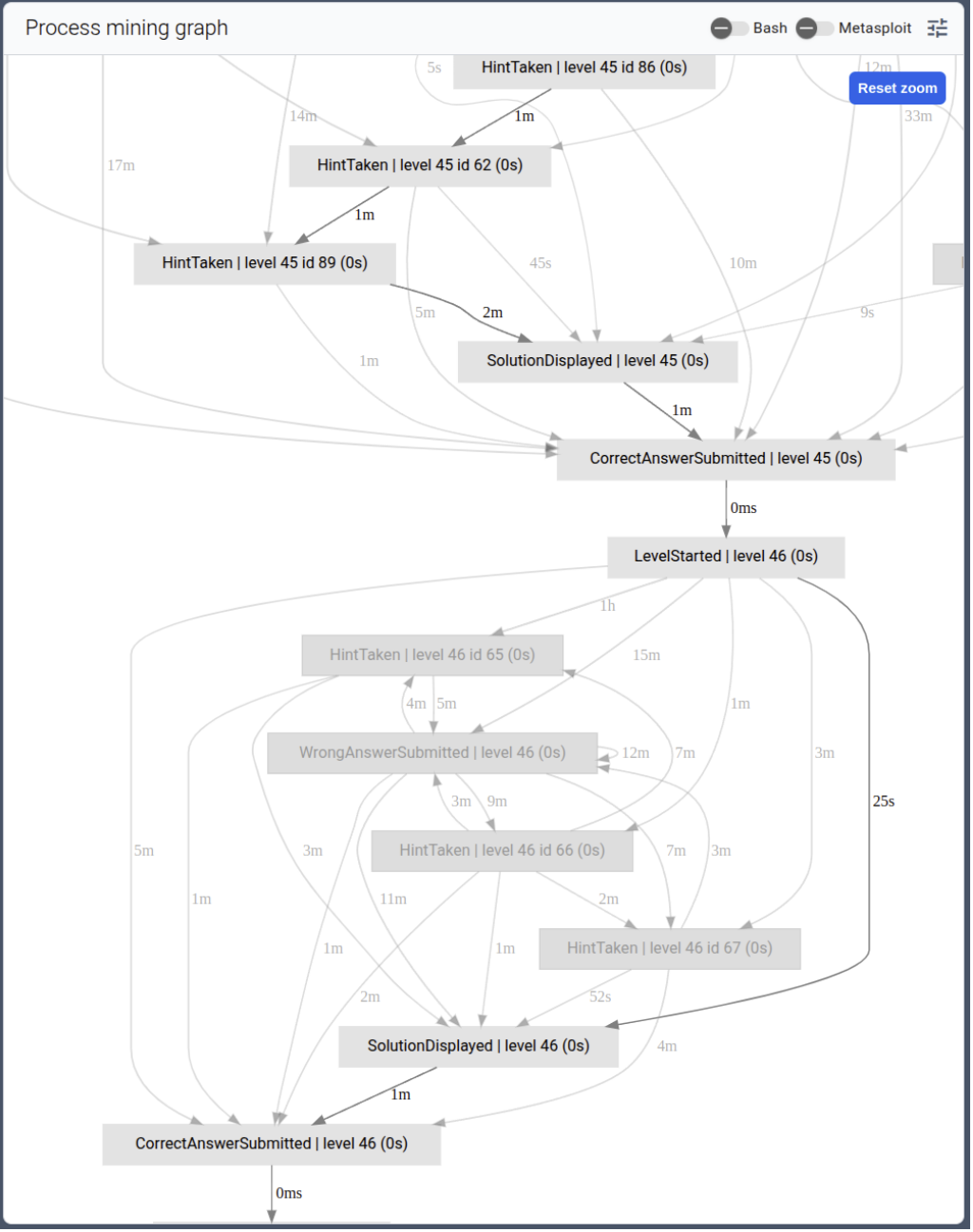}
  \caption{Performance graph of the trainee with low engagement.}
  \label{fig:cause-analysis-perf}
\end{figure*}

The performance graph can reveal a trainee's performance (objective \emph{O2}). Figure~\ref{fig:cause-analysis-perf} displays traces of the selected trainee together with other players (visually suppressed paths). It can be seen that it takes several minutes on average to provide the correct solution to the cybersecurity task (paths between \emph{LevelStarted} and \emph{CorrectAnswerSubmitted} nodes), while the trainee with low engagement went through the game levels very quickly.

These observations suggest that the assignment was either completely incomprehensible to the trainee (possibly due to insufficient initial skills) or that they were simply unmotivated to engage with the CFT game. The data were collected from a training session involving cybersecurity students who were expected to have the necessary knowledge. However, since their performance in the exercise was not formally assessed and only attendance was required to pass, this may explain the trainee’s low engagement.

To examine our expectations about using the dashboard for drill-down investigation of individual behavior, we asked participants of the preliminary evaluation to prove or disprove hypotheses that the low engagement of the selected trainee was caused by the lack of time (\emph{T3}) or insufficient prior knowledge (\emph{T4}). The observation gained from the evaluation and feedback provided by participants has shown that these tasks were more difficult than just detecting trainees with low engagement. Moreover, the analysts were able to confirm their hypothesis with only limited certainty. All respondents came to the conclusion that the primary reason was insufficient prior knowledge of the trainee. Regarding the possible lack of time, the answers covered the full range of possible responses, from ``very unlikely'' to ``very likely''. It is to be pointed out that the respondents were not familiar with the motivation of the assessment context described above.

The drill-down strategies and then the utilization of views and filtering capabilities differed among the participants as well. Some of them responded quickly, relying only on the performance process graph, while others engaged with multiple visualizations and provided more in-depth explanations for their decisions. The sentiment chart, the raw data filtering, and the performance process graph were reported to be the most valuable for cause analysis, followed by the temporary visual suppression, the detailed activity matrix, and the frequency process graph.

\subsection{Lessons Learned From the Evaluation}

Preliminary qualitative analysis conducted on the dashboard's prototype proved the usability of the process-oriented concepts for post-training analysis of cybersecurity of CTF games. On the other hand, the study also identified a few bottlenecks and possible improvements.
\begin{itemize}
    \item The semantics behind the sentiment chart were unclear for analysts who were unsure whether the graph displays trends or specific values.
    \item The performance process graph was primarily used by analysts to answer time-related questions about trainees' walkthroughs. However, some analysts were missing statistical information like that depicted in Figure~\ref{fig:graph-overview} that was not yet implemented in the prototype. One participant mentioned that \emph{``From the performance process graph, I can gain a time-based context regarding the participant's walkthrough. For example, I can observe that the trainee spent 5 minutes on this level, which initially seems reasonable, as they spent a similar amount of time on other levels. However, if I were able to view statistical data for other trainees and found that the average time for this level is only one minute, it would no longer seem fine, and I would want to investigate the issue further.''}
    \item One participant suggested adding icons to the matrix of individual walkthroughs (Figure~\ref{fig:dashboard}(C)) to visually encode types of events and commands, as the current use of colors does not provide any new meaningful information, and the labels are excessively long.
    \item The timeline is crucial for temporal filtering. However, analysts often overlook the current implementation of the timeline in the sentiment chart. A more prominent location and a visually significant redesign are needed.
\end{itemize}

\section{Conclusion and Future Work}

This paper introduces a vital application domain where process mining enhanced with complementary visualizations can help gain insight into the behavior of trainees and possible flaws of hands-on cybersecurity exercises. We addressed three research questions as follows:

\emph{RQ1 -- What are the properties of process graphs derived from event logs of supervised CTF games?} In-depth analysis of real datasets has shown high variability in the size of process graphs. Nevertheless, other revealed features like homogeneity of edge attributes, k-partite structuring of graphs, or significant topological structures allow the designers of analytical tools to select and combine visual approaches respecting the guidelines proposed for the visualizations of multivariate networks.

\emph{RQ2 -- What are the objectives of learning analytics that process mining can help overcome? } Together with domain experts -- practitioners in designing and supervising practical cybersecurity exercises, we formulated three low-level objectives \emph{O1--O3} and three typical post-training questions -- engagement, cheating, and difficulty that can be solved by combining \emph{O1--O3} approaches to prove or disprove corresponding hypotheses. We demonstrated the interactions in the case study of engagement analysis. Also, the results and lessons learned from preliminary user testing were discussed. 

\emph{RQ3 -- What design principles should guide the development of a process-driven analytical dashboard?} Our approach is built upon the conceptual framework for the visual analysis process, with three specific data models being involved: process graphs, models of temporal changes based on sliding windows, and clustering models. Design decisions for visual interactions are backed by the observed properties of process graphs and guidelines formulated for the visualization of multivariate networks. A prototype dashboard described in the paper will serve as a framework for developing and enhancing interactive visualizations in cybersecurity learning analytics.

We believe there are several directions for future work that can significantly strengthen our abilities to investigate practical cybersecurity exercises using human-in-the-loop learning analytics. These research challenges include:
\begin{itemize}
    \item \emph{Other types of cybersecurity games.} As this paper only focused on CTF games, the utilization of process mining and visual analytics in other types of games needs to be explored. In particular, unsupervised training and free cyber defense exercises pose a challenge due to the variability and complexity of game playthroughs, as described in Section~\ref{sec:cyber-training-background}.
    \item \emph{Automatic hints.} Enhancing the proposed approach with automatic hints for the analyst might provide an improved analysis experience. The hints can arise in general uncommon situations, e.g., loops in the process, a very fast transition between game events, and temporal patterns, which can be discovered using a declarative process mining approach.
    \item \emph{Categorization of events.} The current approach can also be enhanced with new perspectives. For example, adding a finer-grained categorization of events to this approach might help the analyst get more abstraction levels and then finely tackle information complexity.
    \item \emph{Real-time analysis.} To enhance situational awareness during the training, there is a possibility to utilize real-time process mining and visualization approaches. Real-time analysis can help to see the current status of the trainees. Furthermore, we can also analyze the probabilities of the training success for each trainee.
    \item \emph{User-defined metrics.} Temporal analysis based on a sliding window can be applied to variable data types, aiming to reflect different behavioral aspects than engagement. The generalization of the principles suggested in this paper and the evaluation of possible metrics is necessary.
    \item \emph{Multi-level abstraction of process graphs.} Although the visual complexity reduction techniques shown in Figure~\ref{fig:graph-overview} had not yet been implemented in the dashboard during the evaluation, one participant identified statistical overviews above the process graph as potentially highly valuable. Therefore, replacing traditional graph-based visualizations with simplified statistical overviews for high-level abstractions could improve drill-down exploration. However, further research is needed to develop suitable visualization techniques.

\end{itemize}

 \bibliographystyle{elsarticle-num} 
 \bibliography{bibliography}

\end{document}